\documentclass[11pt]{article}
\usepackage[toc,page]{appendix}
\usepackage{amssymb}
\usepackage{epsfig}
\begin{document}
%%%%%%%%%%%%%%%%%%%%%%%%%%%%%%%%%
%%%%%%%%%%%%%%%%%%%%
\title{{\bf{\Large Two dimensional hydrodynamics with gauge and gravitational anomalies}}}
%%%%%%%%%%%%%%%%%%%%
\author{  {\bf {\normalsize Rabin Banerjee$^a$}$
$\thanks{E-mail: rabin@bose.res.in}},$~$ 
{\bf {\normalsize Shirsendu Dey$^a$}$
$\thanks{E-mail: shirsendu12@bose.res.in}},$~$
 {\bf {\normalsize Bibhas Ranjan Majhi$^{b,c}$}$
$\thanks{E-mail: bibhas.majhi@mail.huji.ac.il}},
\\$~$
{\bf {\normalsize Arpan Krishna Mitra$^a$}$
$\thanks{E-mail: arpan@bose.res.in}}$~$
\\{\normalsize $^a$S. N. Bose National Centre for Basic Sciences,}
\\{\normalsize JD Block, Sector III, Salt Lake, Kolkata-700098, India}
\\\\
{\normalsize $^b$IUCAA, Post Bag 4, Ganeshkhind,}
\\{\normalsize Pune University Campus, Pune 411 007, India}
\\
{\normalsize $^c$Racah Institute of Physics, Hebrew University of Jerusalem,}
\\{\normalsize Givat Ram, Jerusalem 91904, Israel}
\\[0.3cm]
} 
%%%%%%%%%%%%%%%%%%%%%%%%%%%%%%%%%%%%%%%%%%
%\date{}

\maketitle

\begin{abstract}
We present a new approach to discuss two dimensional chiral and non-chiral hydrodynamics with gauge and 
gravitational anomalies. Exact constitutive relations for the stress tensor and 
charge current are obtained. For the chiral theory, the constitutive relations may be put in the ideal ($chiral$) fluid form whereas the constitutive relations corresponding to non-chiral case do not take the ideal fluid form. The constitutive relations in the presence of both gravity and gauge sectors are new. These expressions, in the absence of the gauge sector, reproduce the results obtained in the gradient expansion approach. 
\end{abstract}
%%%%%%%%%%%%%%%%%%%%%%%%%%%%%%%%%%%%%%%%%%%%%%%%%%%%%%%%%%%%%%
\section{\label{intro}Introduction}

  Hydrodynamics is basically a finite temperature quantum field theory in the limit of large time and length scales \cite{Landau}. Certain conservation laws govern it which are manifested by the global symmetries of the underlying theory. For instance, the conservation of the energy-momentum tensor invokes the spacetime symmetry while the conservation of charge current is yielded by the charge symmetry. The
 energy-momentum tensor and the U(1) current have to satisfy the
 conservation equations. The constitutive equations give an explicit
 form for the one-point functions of the energy-momentum tensor and the U(1) current in terms of the hydrodynamic variables e.g. velocity, temperature, chemical potential, etc. These constitutive equations are traditionally required to make the local production of entropy positive, in accordance with a local version of the second law of thermodynamics.

 Recently, investigations on fluid dynamics in the presence of anomalies have received considerable attention \cite{Rangamani:2009}-\cite{Banerjee:2013qha}. This is important because one expects that the standard picture might non-trivially be modified in the presence of the quantum anomalies.
In this paper we like to analyse the structure of constitutive relations in the presence of both gauge and gravitational
anomalies. The later includes diffeomorphism anomaly (violation of general coordinate invariance) as well as trace anomaly (violation of conformal invariance). The present analysis is confined to ($1+1$) dimensions. Incidentally two dimensions have certain peculiarities which make it feasible to abstract various results that may not be possible in higher dimensions. 
It is well known that the two dimensional metric, in general, can be expressed in a conformally flat 
form. Therefore, the effective action for a field under this background is exactly obtainable and one does not
need the derivative expansion method. This ensures the transparency  of the physics of the problem at each stage of the computation. 

 In this paper, we discuss both the non-chiral and chiral theories. At the classical level there are no anomalies. For a quantum theory a one loop computation leads to anomalies. In the case of a non-chiral theory (i.e. where ordinary fermions are coupled to gauge and/or gravity sector) the anomaly is a manifestation of the clash between general coordinate invariance and conformal invariance. For a gravitational theory, usually the former is retained at the expense of the later. Hence one has to appropriately account for the conformal (trace) anomaly. A chiral theory, on the other hand, has both diffeomorphism and conformal anomalies, since it is not possible to preserve any of the symmetries at the quantum level.

 We develop a new method to discuss anomalous hydrodynamics in two dimensions that includes both gauge and gravity sectors. We start from the anomalous energy-momentum tensor and current obtained from the corresponding effective action and then the components are evaluated explicitly for a general static metric in null coordinates. Solutions of 
  the anomalous expressions involve integration constants. These constants crop up in the various expressions for the constitutive relations, which are found by introducing the appropriate fluid variables, for the stress tensor and the gauge current. The constitutive relations found here are exact and do not require any gradient expansion as happens in higher dimensions. These relations are new and have not been posited earlier in the literature \cite{Jensen:2012jh}-\cite{Jains:2013}. Moreover, for the chiral theory, the constitutive relations may be put in a form that resembles the structure for an $ideal~ chiral$ fluid. The constitutive relation for an $ideal~ chiral$ fluid, it may be recalled, has a form similar to the usual ideal fluid \cite{Landau} but with the velocity vector replaced by the chiral velocity vector \cite{Banerjee:2013qha}. It is reassuring to note that our results agree with those found by the derivative expansion approach when the gauge sector is absent\cite{Jensen:2012jh}. Incidentally, in the presence of this sector, results are not available in the gradient expansion method.

   The organization of our paper is as follows. A general setup for our main purpose 
is presented in the next section.  In section \ref{review}, we provide a brief discussion, 
some of which may not be that well known, on two dimensional gauge and gravitational anomalies. 
Then we find the constitutive relations for the non-chiral case in section \ref{vector}. 
The chiral case is analysed in the next section. Section \ref{comparison} is devoted to a comparison with the 
gradient expansion approach. The constitutive relations obtained by us contained an arbitrary constant. For a particular value of this constant our expressions (in the absence of a gauge field) reproduce those found by the derivative expansion method. Inclusion of gauge fields has not been considered in full in the derivative expansion method, hence a comparison with our result is not feasible.  Finally, we conclude in section \ref{conclu}. We also added an Appendix. 
 %%%%%%%%%%%%%%%%%%%%%%%%%
 \section{\label{metric}Metric and general setup}
   Here we shall consider a ($1+1$) general static spacetime. Also the explicit components of the comoving velocity vector will be calculated under this background, which is necessary for our main purpose. All the required expressions will sometimes be expressed in null coordinates to make them more transparent.

 The static metric in ($1+1$) dimensions can be taken in the following form:
\begin{equation}
ds^2 = - e^{2\sigma(r)}dt^2 + g_{11} dr^2~.
\label{effectivemetric}
\end{equation}
It has a timelike Killing vector and the Killing horizon is given by the solution of the equation $e^{2\sigma(r)}|_{r_0} = 0$. The $U(1)$ gauge fields $A_a$ are of the linear form
\begin{equation}
A_a = (A_t(r),0)~.
\label{gauge}
\end{equation}
The null coordinates ($u,v$) are defined as
\begin{equation}
u=t-r_*; \,\,\,\ v=t+r_*~,
\label{null}
\end{equation}
where the tortoise coordinate $r_*$ is given by $dr_* = - \sqrt{g_{11}}~ e^{-\sigma}dr$. In these coordinates, the metric takes the following off-diagonal form:
\begin{equation}
ds^2 = - \frac{e^{2\sigma}}{2} (dudv + dvdu)~.
\label{nullmetric}
\end{equation}
In order to express the anomalous energy-momentum tensor in terms of fluid variables, we adopt the comoving frame. In this frame, the velocity vector of a fluid $u^a$ is normalised as,
\begin{equation} 
u^au_a=-1.
\label{velnor}
\end{equation}
Note that the norm of the velocity has to be negative since we are considering time like trajectories. The absolute normalisation is fixed to unity by choosing the comoving frame. Subjected to the above condition and the metric (\ref{effectivemetric}), the usual ansatz for the velocity follows,
\begin{equation}
u^a = e^{-\sigma}(1,0); \,\,\,\ u_a = -e^{\sigma}(1,0)~,
\label{velocity}
\end{equation}
with $a=t,r$. 
Correspondingly, in null coordinates ($u,v$), this transforms to the following form: 
\begin{equation}
u^a = e^{-\sigma}(1,1); \,\,\,\ u_a = -\frac{e^{\sigma}}{2}(1,1)~.
\label{velonull}
\end{equation}
With the above expressions for the velocity vector, the following combinations of fluid variables can be explicitly expressed in terms of the metric coefficients:
\begin{equation}
u^a\nabla^b\nabla_a u_b = \frac{1}{2g^2_{11}}(2g_{11}\sigma'' + 2 g_{11}\sigma'^2 - \sigma'g'_{11}); \,\,\,\,\ u^a\nabla^b\nabla_bu_a  = \frac{\sigma'^2}{g_{11}}~,
\label{fluidvar}
\end{equation}
which are necessary to find the constitutive relations.

  % Before proceeding further, we mention that in two dimensions the metric will have three independent metric coefficients. Since we are interested in the static case, the most general form can be taken as \cite{Jains:2013}
%\begin{equation}
%ds^2 = - e^{2\sigma(r)}(dt + a_1(r)dr)^2 + g_{11}(r) dr^2~.
%\label{generalmetric}
%\end{equation}
%Now it is always possible to reduce the degrees of freedom by exploiting coordinate transformations. This can be done here by using the transformation $dt\rightarrow dt- a_1(r)dr $. It will immediately reduce the above metric to the form given earlier in (\ref{effectivemetric}).
%Therefore the fluid velocity, considered in (\ref{velocity}), is the most general form in comoving frame (in two dimension) in the absence of any dissipation, for the metric (\ref{effectivemetric}).    

  To discuss the ($1+1$) dimensional chiral theory, in which both the diffeomorphism and trace anomalies appear, we define chiral velocity as
\begin{equation}
{u^{(c)}}_a=u_a-{\tilde{u}}_a
\label{chiralvel}
\end{equation}
where $ \tilde{u}_a = {\bar{\epsilon}}_{ab}u^b$ is the dual to $u_a$ and $\bar{\epsilon}_{ab}$ is an antisymmetric tensor with $\bar{\epsilon}_{ab} = \sqrt{-g}\epsilon_{ab}$ and $\bar{\epsilon}^{ab} = \epsilon^{ab}/\sqrt{-g}$.
In null coordinates the components are given by,
\begin{equation}
\label{vel}
\epsilon_{uv} = 1; \,\,\ \epsilon^{uv}=-1; \,\,\ {u^{(c)}}_a=-e^\sigma{(1,0)}.
\end{equation}
Note that the definition (\ref{chiralvel}) of the chiral velocity ${u^{(c)}}_a$  ensures that it satisfies the familiar chiral property in two dimensions,
\begin{equation}
{u^{(c)}}_a=-{\bar{\epsilon}}_{ab}{u^{(c)}}^{b}
\label{chiral}
\end{equation}
%%
%%%%%%%%%%%%%%%%%%%%%%%%%%%%%%%%%%%%%%%%%%%%%%%%%%%%%
%%%%%%%%%%%%%%%%%%%%%%%%%%%%%%%%%%%%%%%%%%%%%%%%%%%%%%%%%%
 \section{\label{review}Review of gauge and gravitational anomalies}
  An anomaly is a breakdown of some classical symmetry upon quantization. It may have different manifestations but generally speaking these are connected. A violation of gauge symmetry is revealed by a non-conservation of the gauge current (gauge anomaly) or, alternatively, by the presence of anomalous terms in the algebra of currents. These anomalous terms are related to the gauge anomaly. Likewise , a violation of diffeomorphism symmetry leads to the non-conservation of the stress tensor.

  Of particular significance are theories where chiral symmetries are gauged. The equations of motions show that the chiral currents are covariantly conserved. However, quantum effects destroy this feature as may be checked by doing a one loop computation. Algebraically, this may be expressed as,
\begin{eqnarray}
{\left\langle {D_\mu J^\mu } \right\rangle}^a = \partial_\mu  {\left\langle J^{\mu a}\right\rangle} - f^{abc} A_\mu^b {\left\langle J^{\mu c}\right\rangle} = G^a
\label{revavg}
\end{eqnarray}
where the average is interpreted to be taken over the fermionic degrees of freedom appearing in the chiral current  $ J^a_\mu $ and the other symbols have their usual meaning. $G^a$ is called the anomaly.

  There are different ways of defining the averaged current corresponding to different regularisation choices. Among the various possibilities two are outstanding. The first and perhaps the more common way is to interpret the averaged current as the functional derivative of an effective action $W$,
\begin{eqnarray}
{\left\langle J_\mu ^a{(x)}\right\rangle}= \frac{\delta W}{\delta A^{a\mu}{(x)}}
\label{revW}
\end{eqnarray}
This current satisfies the integrability condition,
\begin{eqnarray}
 \frac{\delta J^{\mu a}{(x)}}{\delta A^b_\nu{(x')}}= \frac{\delta J^{\nu b}{(x')}}{\delta A^a_\mu{(x)}}
\label{revinte}
\end{eqnarray}
A direct consequence of this relation is that the anomaly of this current satisfies the Wess-Zumino consistency condition. To obtain this condition let us introduce the operator ,
\begin{eqnarray}
L^a{(x)}~=~ \partial_\mu \frac{\delta}{\delta A^a_\mu{(x)}}- f^{abc} A^b_\mu{(x)}\frac{\delta}{\delta A^c_\mu{(x)}}
\label{revop}
\end{eqnarray}
which is basically the generator of the infinitesimal gauge transformation. From (\ref{revavg}), (\ref{revW}) and (\ref{revop}) it follows that,
\begin{eqnarray}
L^a{(x)}W~=~G^a{(x)}
\label{revgen}
\end{eqnarray}
so that the existence of the anomaly is a statement about the lack of gauge invariance of the one loop effective action. The generators satisfy the closed algebra,
\begin{eqnarray}
\left[L^a{(x)}, L^b{(y)}\right]~= f^{abc} \delta{(x-y)} L^c{(x)}
\label{revcomm}
\end{eqnarray}
Acting on $W$ and using (\ref{revgen}) immediately yields,
\begin{eqnarray}
L^a{(x)}G^b{(y)}-L^b{(y)}G^a{(x)}=~f^{abc} \delta{(x-y)} G^c{(x)}
\label{revLG}
\end{eqnarray}
This is the Wess-Zumino condition. It is now clear that the anomaly of the current (\ref{revW}) defined through the effective action must satisfy this consistency condition and is hence called the consistent anomaly.

   The other way of defining the averaged current is to regularise it by a gauge covariant method. In that case the current transforms covariantly under the gauge transformation so that,
\begin{eqnarray}
L^a{(x)} J_\mu^b{(y)}~=~f^{abc} \delta{(x-y)}J^c_\mu
\label{revLJ}
\end{eqnarray}
It is then simple to show, by taking covariant divergence of both sides of this equation, that the corresponding anomaly $G^a$ also transforms covariantly,
\begin{eqnarray}
L^a{(x)} G^b_{cov}{(y)}~=~f^{abc} \delta{(x-y)}G^c_{cov}
\label{revLG1}
\end{eqnarray}
This anomaly is called the covariant anomaly. It is easy to see, by an appropriate change of indices, that this anomaly satisfies the condition,
\begin{eqnarray}
L^a{(x)}G^b_{cov}{(y)}-L^b{(y)}G^a_{cov}{(x)}=~2f^{abc} \delta{(x-y)} G^c_{cov}{(x)}
\label{revLGcov}
\end{eqnarray}
Comparison with (\ref{revLG}) immediately shows that the covariant anomaly is incompatible with the Wess-Zumino condition, it is off by a factor of 2. This analysis illustrates the difference between covariant and consistent expressions. While the covariant anomaly transforms covariantly under a gauge transformation but does not satisfy the Wess-Zumino condition, the behaviour of the consistent anomaly is just the reverse. It satisfies the Wess-Zumino condition but does not transform covariantly. Since currents and/or stress tensors are only defined modulo local counter terms manifesting the regularisation ambiguities, covariant and consistent expressions are also related by such counter terms. These local polynomials were obtained by using either differential geometric methods \cite{Bardeen:1984pm} or by dynamical means \cite{Banerjee:1985ti,Banerjee:1986bu}.

  The above discussion is simply illuminated by means of the two dimensional example which is the case considered here. Using a covariant regularisation, the covariant (gauge) anomaly is found as {\footnote{A word about the notation. Space-time indices are denoted in this section by Greek letters $\mu$, $\nu$ etc. In other sections it is denoted by Latin $a,b$ etc. Here $a,b$ stand for the non-abelian group indices.}}
\begin{eqnarray}
&& G_{cov} = \frac{1}{4\pi} \epsilon_{\mu\nu}F ^{\mu\nu}
\label{revGcov}
\end{eqnarray}
where we have considered the gauge group to be abelian.

The Euclidean effective action is defined as \cite{Banerjee:1985ti,Banerjee:1986bu},
\begin{eqnarray}
&& W~=~\int^1_0 dg~\int d^{2}x~A_{\mu}\left(x\right) J^{(g)}_{\mu}(x) 
\label{reveffaction}
\end{eqnarray}
where the superscript $``g"$ indicates that this is the coupling constant to be used in the construction of the current $J_{\mu}(x)$. The above equation is a formal definition of $W$. To concretise it, one has to regularise the current. Let us regularise it covariantly. Then, under an abelian gauge transformation,~$A_{\mu}\rightarrow~A_{\mu}-\partial_{\mu}\alpha$,
\begin{eqnarray}
&& \int d^{2}x~\left\lbrace \partial_{\mu} \frac{\delta W}{\delta A_{\mu}}\right\rbrace \alpha = \int_0^1 dg \int d^{2}x~\alpha~ \partial_{\mu}J^{\left( g\right) }_{\mu}(x)
\label{revreg}
\end{eqnarray}
Since $ \alpha $ is arbitrary, equating the integrands yields,
\begin{equation}
\label{re} 
\partial_{\mu} \frac{\delta W}{\delta A_{\mu}}=\int^{1}_{0} dg ~\partial_{\mu} J^{(g)}_{\mu}(x).
\end{equation}
Inserting the value of the covariant anomaly (\ref{revGcov}) yields the consistent anomaly,
\begin{eqnarray}
&& \partial_{\mu} \frac{\delta W}{\delta A_{\mu}}~=~\frac{1}{8\pi} \epsilon_{\mu \nu} F^{\mu \nu}
\label{revcons}
\end{eqnarray}
where the half factor comes from the integration over $``g"$ since it occurs linearly in $ F_{\mu \nu}(=g(\partial_{\mu}A_{\nu}-\partial_{\nu}A_{\mu}))$. Indeed, in any arbitrary even $d=2n$ dimensions the (abelian) consistent and covariant anomalies are related as, 
\begin{eqnarray}
G_{cons} = \frac{1}{n+1} G_{cov}
\label{revGcons}
\end{eqnarray}
since $F$ involves $g$ homogeneously to the $n^{th}$ power.
      It is now straightforward to check that the local polynomial connecting the consistent and covariant currents is given by,
\begin{eqnarray}
J^{cons}_{\mu} = \frac{\delta W}{\delta A_{\mu}} = J^{cov}_{\mu} - \frac{1}{4\pi}\epsilon_{\mu \nu} A_{\nu} 
\label{revJcons}
\end{eqnarray}
such that compatibility with (\ref{revGcov}) and (\ref{revcons}) is established.

   We now make an important point. Initially (\ref{reveffaction}) was defined by taking $J_{\mu}$ to be covariant. If we take $J_{\mu}$ to be consistent then the result remains unaffected since the difference vanishes,
\begin{eqnarray}
\int^1_0 dg \int d^{2}x ~A_{\mu}(\epsilon_{\mu \nu} A_{\nu})=0
\label{revzero}
\end{eqnarray}
This shows that the effective action $W$ remains the same whether the current is regularised covariantly or consistently. It is a general result valid in any dimensions \cite{Banerjee:1985ti,Banerjee:1986bu}.

Let us next consider gravitational anomalies. Such anomalies occur in $4n+2$ dimensions $(n=0,1,2...)$ in contrast to gauge anomalies that occur in $2n$ dimensions $(n=1,2...)$\cite{Gaume-Gins:1985,Gaume-witten:1984}. This shows that two dimensions are slightly special. It is the simplest space-time dimension where both gauge and gravitational anomalies may be present. Since  our subsequent analysis will be done for two dimensions we will restrict our discussion on gravitational anomalies also for this case. Although the results are known, this presentation is given primarily for two reasons. First, the results are obtained in a simple and elementary way using special properties of two dimensions. Secondly, this approach, discussed previously \cite{Banerjee:2008sn,Banerjee:2008wq} in 
fragmented parts, is not particularly well known.

We shall derive the result  for the two dimensional  gravitational (diffeomorphism) anomaly that is known to exist in a chiral theory. The anomaly will be obtained directly in the covariant form which will be exploited in later sections to derive the constitutive relations. As mentioned previously, different manifestation  of anomalies are related. Here we show the obtention of the diffeomorphism anomaly in a chiral theory from the conformal (trace) anomaly in a non-chiral (vector-like) theory. In the later theory it is well known that it is not possible , at the quantum level, to simultaneously preserve both diffeomorohism  and conformal symmetries that are present classically. Since  diffeomorphism invariance is regarded as more fundamental in a gravitational theory, this is retained at the quantum level at the expense of conformal invariance. The breakdown of conformal invariance leads to  the  trace anomaly,
\begin{eqnarray}
\label{rea}
T^{\mu} _{\mu}= \frac{R}{24\pi}
\end{eqnarray} 
where $R$ is the Ricci scalar.

Now the energy momentum tensor in two dimensions can decomposed into a traceful and traceless part as,
\begin{equation}
\label{reb}
T_{\mu\nu}= \frac{R}{48\pi}g_{\mu\nu}+\theta_{\mu\nu}
\end{equation} 
where $\theta_{\mu\nu}$ is symmetric $(\theta_{\mu\nu}= \theta_{\nu\mu})$ to preserve the symmetric property of $T_{\mu\nu}$ and traceless $\theta^{\mu}~_{\mu}=0$. Taking the trace of (\ref{reb}) then yields (\ref{rea}). Furthermore, since general coordinate invariance is preserved $( \bigtriangledown^{\mu} T_{\mu\nu}=0 )$, it implies the following constraint on $\theta_{\mu\nu}$,
\begin{equation}
\label{rec}
\bigtriangledown^{\mu} \theta_{\mu\nu}= - \frac{1}{48\pi} \bigtriangledown_{\nu}R
\end{equation}
The stress tensor (\ref{reb}) may be interpreted as the sum of the contributions from the right and left moving modes. Moreover, the      left-right symmetry implies that the contribution from one mode is equal to that from the other mode, except that the $u$ and $v$ variables have to be interchanged. Since $T_{\mu\nu}$ is symmetric we have $T_{\mu\nu}= T^{(R)}_{\mu\nu}+T^{(L)}_{\mu\nu}$ with, 
\begin{equation}
\label{red}
T^{R(L)}_{\mu\nu}= \frac{R}{96\pi}g_{\mu\nu}+\theta^{R(L)}_{\mu\nu}
\end{equation}
where $\theta_{\mu\nu}= \theta^{(R)}_{\mu\nu}+\theta^{(L)}_{\mu\nu}$ (in analogy with $T_{\mu\nu}$). The chirality (or holomorphy) condition on $T^{R(L)}_{\mu\nu}$ implies the following equality \cite{Banerjee:2008sn,Banerjee:2008wq},
\begin{equation}
\label{ree}
T^{R(L)}_{\mu\nu}= -(+) \frac{1}{2}(\bar \epsilon_{\mu\sigma}T~_{\nu}^{\sigma ~R(L)}+\bar \epsilon_{\nu\sigma}T~_{\mu}^{\sigma ~R(L)})\\
+ \frac{1}{2}g_{\mu\nu} T~_{\alpha}^{\alpha~ R(L)}
\end{equation}
Hence we have the following equations for the right and left modes;
\begin{equation}
\label{ref}
T^{R}_{vv} = T^{L}_{uu}=~0,~T^{R}_{uu}=T^{L}_{vv}\neq 0
\end{equation}
Expectedly, the $ L \leftrightarrow R $ symmetry under the interchange $ u \leftrightarrow v $ is preserved. The above conditions along with the tracelessness of $ \theta_{\mu\nu} $ yield further relations that follow from (\ref{red}),
$$\theta^{R}_{uv}=\theta^{R}_{vv}=0,~ \theta^{R}_{uu}\neq 0$$
\begin{equation}
\label{reg}
\theta^{L}_{uv}=\theta^{L}_{uu}=0,~ \theta^{L}_{vv}\neq 0
\end{equation}
With this input it is possible to deduce the various anomalies. The trace anomaly for the chiral theory is quickly  obtained from (\ref{red})
\begin{equation}
\label{reh}
T^{\mu~(R)}_{\mu}=T^{\mu~(L)}_{\mu}=\frac{R}{48\pi}
\end{equation}
which is half the result for the usual theory. It needs a little bit more algebra to derive  the diffeomorphism anomaly  for the chiral components. Taking the right mode and applying  the covariant derivative on (\ref{red}) yields, 
\begin{equation}
\label{rei}
\nabla^{\mu} T^{R}_{\mu\nu}= \frac{1}{96\pi}\nabla_{\nu}R + \nabla^{\mu} \theta^{R}_{\mu\nu} 
\end{equation}
Next, using (\ref{rec}) and (\ref{reg}) for the R-mode, we find,
\begin{equation}
\label{rej}
\nabla^{\mu} \theta^{R}_{\mu u}= \frac{1}{48\pi}\nabla_{u}R ~,~ \nabla^{\mu} \theta^{R}_{\mu v}=0 
\end{equation}
Inserting these expressions in (\ref{rei}) we obtain for $\nu= u$ and $\nu=v$,
\begin{eqnarray}
\label{rek}
\nabla^{\mu} T^{R}_{\mu u}= \frac{1}{96\pi}\nabla_{u}R - \frac{1}{48\pi}\nabla_{u}R = -\frac{1}{96\pi}\nabla_{u}R
\end{eqnarray}
and,
\begin{eqnarray}              
\nabla^\mu   T^{R}_{\mu v}=\frac{1}{96\pi}\nabla_{v}R
\end{eqnarray}
Combining them yields,
\begin{equation}
\label{rel}
\nabla^{\mu} T^{R}_{\mu\nu}= \frac{1}{96\pi}  \bar \epsilon_{\nu\sigma} \nabla^{\sigma}R
\end{equation}
which is the cherished covariant gravitational anomaly. One can repeat the calculation for the left mode or directly obtain the final result by using the $ L\leftrightarrow R$ symmetry under $u \leftrightarrow v$. The result is same as (\ref{rel}) except for a change in sign. Since covariant expressions have been used throughout, the final result is also covariant. It is possible to obtain the consistent anomaly by adding local polynomials. However this issue need not concern us since we will be dealing with the covariant anomaly only.

~To summarise, whereas the non-chiral theory has a trace anomaly (\ref{rea}) but no diffeomorphism anomaly, the chiral theory admits both types of anomalies (\ref{reh},\ref{rel}). Physically, this is related to the unidirectional property of a chiral theory \cite{Fulling:1986rk,Banerjee:2008wq}. This distinction has an important role in the structure of the constitutive relations.
%%%%%%%%%%%%%%%%%%%%%%%%%%%%%%%%%%%%%%%%%%%%%%%%%%%
\section{\label{vector}Anomalous constitutive relations in hydrodynamics}
%%%%%%%%%%%%%%%%%%% 

~As elaborated in the previous section, for the non-chiral theory, after quantization, either the trace or the diffeomorphism anomaly exists. Usually, one likes to 
retain the diffeomorphism symmetry at the cost of the conformal symmetry. In that case, one has the 
trace anomaly. In ($1+1$) dimension the corresponding effective action is given by 
the Polyakov form \cite{Polyakov:1981rd}: 
\begin{eqnarray}
&&S_P^{(g)} = \frac{1}{96\pi}\int d^2xd^2y\sqrt{-{{g}}} {{R(x)}}\frac{1}{{\Box}}(x,y)\sqrt{-g}{{R}}(y)~;
\label{effective1}
\\
&&S_P^{U(1)} = \frac{e^2}{2\pi} \int d^2xd^2y\sqrt{-{{g}}} {\bar{\epsilon}}^{ab}\partial_aA_b(x)\frac{1}{\Box}(x,y)\sqrt{-g}{\bar{\epsilon}}^{cd}\partial_cA_d(y)~,
\label{effectiveU1}
\end{eqnarray} 
where $S_P^{(g)}$ is the effective action for the gravity sector whereas $S_P^{U(1)}$ is that for the gauge sector. 
The total action is $S_P = S_P^{(g)} + S_P^{U(1)}$.
$\frac{1}{{{\Box}}}$ is the inverse of d'Alembertian ${\Box} = {\nabla}^a {\nabla}_a = \frac{1}{\sqrt{-g}} \partial_a \left(\sqrt{-g}~g^{ab}{\partial}_b \right)~$.
This action is non-local but it can be written in a local form by introducing auxiliary fields ${\Phi}$ and $B$, defined as
\begin{eqnarray}
&&\Phi(x) = \int d^2y \frac{1}{\Box}(x,y)\sqrt{-g}R(y)~; 
\label{auxphi}
\\
&&B(x)=\int d^2y \frac{1}{\Box}(x,y)\sqrt{-g}{\bar{\epsilon}}^{ab}\partial_aA_b(y)~. 
\label{uxB}
\end{eqnarray}
Then,
\begin{eqnarray}
&&S_P^{(g)} = \frac{1}{96\pi}\int d^2x{\sqrt{-{g}}} (-{\Phi} {\Box} {\Phi} + 2{\Phi} {R})~;
\label{effective2}
\\
&&S_P^{U(1)} = \frac{e^2}{2\pi}\int d^2x\sqrt{-g}(-B\Box B + 2{\bar{\epsilon}}^{ab} \partial_aA_b~B)~.
\label{effectiveU2}
\end{eqnarray}
The above structure of the effective action is the general form for theories where usual (and not chiral) fermions or scalars are coupled to the gauge and gravitational fields. The chiral case is considered in the next section. 
%These forms are strictly valid for the specific theories for which they were meant. It is possible to discuss more general structures, as for example provided in \cite{Jensen:2012kj}, whose specific parametrisations reproduce these forms. 
We prefer to discuss the models, case by case, in order to highlight the interplay between the conformal and diffeomorphism anomalies, apart from generating exact results.

The two dimensional anomalous energy-momentum tensor and the $U(1)$ current are given by, 
\begin{eqnarray}
&&{{T}}_{ab}^{(g)}= -\frac{2}{\sqrt{-{g}}}\frac{\delta S_P^{(g)}}{\delta{g}^{ab}} = \frac{1}{48\pi}\Big[{\nabla}_a{\Phi}{\nabla}_b{\Phi} - 2{\nabla}_a{\nabla}_b{\Phi} + {g}_{ab}\Big(2{R} - \frac{1}{2}{\nabla}_c{\Phi}{\nabla}^c{\Phi}\Big)\Big]~;
\label{tensorg}
\\
&&T_{ab}^{U(1)} =  -\frac{2}{\sqrt{-{g}}}\frac{\delta S_P^{U(1)}}{\delta{g}^{ab}} = \frac{e^2}{\pi}\Big[\nabla_aB\nabla_bB - \frac{1}{2}g_{ab}\nabla^cB\nabla_cB\Big]~,
\label{tensorU}
\end{eqnarray} 
and 
\begin{equation}
J^a =\frac{1}{\sqrt{-g}}\frac{\delta S_p^{U(1)}}{\delta {A}_a}= \frac{e^2}{\pi}{\bar{\epsilon}}^{ab}\partial_b B~,
\label{current}
\end{equation}
respectively.
The auxiliary fields ${\Phi}$ and $B$ satisfy the following equations of motion:
\begin{eqnarray}
{\Box} {\Phi} = {R}~; \,\,\,\ \Box B = \bar{\epsilon}^{ab}\partial_aA_b~.
\label{boxR}
\end{eqnarray}
The total energy-momentum tensor is $T_{ab} = T_{ab}^{(g)} + T_{ab}^{U(1)}$.
Here $R$ is the two dimensional Ricci scalar which, for the metric (\ref{effectivemetric}), is given by
\begin{equation}
R = \frac{1}{g_{11}^2}(g_{11}^{'}\sigma' - 2g_{11}\sigma'^2 - 2g_{11}\sigma^{''})~.
\label{Ricci}
\end{equation}
It turns out that (\ref{tensorg}) leads to $\nabla_a T^{ab(g)} = 0$ and non-vanishing of trace whereas 
(\ref{tensorU}) yields $\nabla_b{T^{ab}}^{U(1)}=J_bF^{ab}$ and vanishing of trace of the stress-tensor. 
Therefore, the explicit form of the trace anomaly of the theory is solely given by the gravity part.
Hence we have
\begin{equation}
T^{a(g)}_{a} = \frac{R}{24\pi}~; \,\,\,\ T^{a~U(1)}_{a} = 0~.
\label{trace}
\end{equation}
Next, we shall evaluate the explicit expressions for the components of the energy-momentum tensor and $U(1)$ current in null coordinates for the background (\ref{nullmetric}).

The ($uv$) component of stress tensor is easy to find because in null coordinates it is proportional to the trace. Therefore from (\ref{trace}) we obtain,
\begin{equation}
T_{uv}^{(g)} = -\frac{e^{2\sigma} R}{96\pi}~; \,\,\,\ T_{uv}^{U(1)} = 0~.
\label{uv}
\end{equation}
To find the other components, we need to start from (\ref{tensorg}) and (\ref{tensorU}) where the auxiliary fields are determined by the solutions of (\ref{boxR}).  
The solutions of (\ref{boxR}), for the background  (\ref{effectivemetric}), turn out to be
\begin{eqnarray}
&&\Phi = \Phi_0(r) - 4pt + q; \,\,\,\ \partial_r\Phi_0 = -2\sigma' + z\sqrt{g_{11}}~ e^{-\sigma}~,
\label{solphi}
\\
&& B = B_0(r) + Pt +Q; \,\,\,\ \partial_r B_0(r) = -e^{-\sigma}\sqrt{g_{11}}(A_t(r) + C)
\label{solB}
\end{eqnarray}
where $p$, $q$, $z$, $P$, $Q$ and $C$ are constants.  A detailed analysis to find the solutions is given in the Appendix.
Then (\ref{tensorg}) and (\ref{tensorU}) yield,
\begin{eqnarray}
&&T_{uu}^{(g)} = \frac{e^{2\sigma}}{96\pi g_{11}^2} (2\sigma''g_{11} - \sigma'g'_{11}) + C_{uu}~;
\label{uu}
\\
&&T_{vv}^{(g)} = \frac{e^{2\sigma}}{96\pi g_{11}^2} (2\sigma''g_{11} - \sigma'g'_{11}) + C_{vv}~;
\label{vv}
\end{eqnarray} 
and 
\begin{eqnarray}
T_{uu}^{U(1)} = \frac{e^2}{4\pi} (A_t-P+C)^2; \,\,\,\ T_{vv}^{U(1)} = \frac{e^2}{4\pi} (A_t+P+C)^2~,
\label{uuU}
\end{eqnarray}
where $C_{uu}$ and $C_{vv}$ are constants, made out of $p$ and $z$.
Similarly, the components  of the current are 
\begin{equation}
J_u = \frac{e^2}{2\pi} (A_t-P+C); \,\,\,\ J_v = \frac{e^2}{2\pi} (A_t+P+C)~.
\label{J}
\end{equation}

We now construct the constitutive relation for the energy momentum tensor of anomalous hydrodynamics. This will be done as follows. In the comoving frame, the fluid velocity components are given in (\ref{velonull}). Using (\ref{velonull}),(\ref{fluidvar}) and (\ref{Ricci}), apart from also using the Tolman relation for the temperature $T= T_0e^{-\sigma}$, where $T_0$ is the equilibrium temperature, and the chemical potential, $\mu = A_t e^{-\sigma}$, the constitutive relations may be determined.
%\footnote{We may recall that for a black hole, the Tolman relation gives the observer dependent temperature T. For an observer at infinity, this T reduces to $T_0$ which is the Hawking temperature.}
 The various components can be written in the following covariant forms:
\begin{eqnarray}
T_{ab}^{(g)}&=&\Big[\frac{1}{12\pi}(u^c\nabla^d -u^d\nabla^c)\nabla_c u_d +4\bar{C}T^2\Big]u_a u_b - \Big[\frac{1}{24\pi} u^c\nabla^d \nabla_d u_c -2 \bar{C}T^2\Big] g_{ab}~;
\label{generalTg}
\\
T_{ab}^{U(1)} &=& \left[\frac{e^2}{2\pi}( \mu^2 +\bar{{C_1}^2}T^2)\right](2u_au_b+g_{ab}) +\left[\frac{e^2}{\pi}\mu \bar{C_1}T\right]\left({u}_a\tilde{u}_b+\tilde{u}_a{u}_b\right)~;
\label{generalTU}
\\
J_a &=& -\frac{e^2}{\pi}\left(\mu+C\frac{T}{T_0}\right) u_a-\left(\frac{e^2 P}{\pi}\right) \frac{T}{T_0}\tilde{u_a}~=-\frac{e^2}{\pi}\left(\mu +\bar{C_1}T\right)u_a,
\label{generalJ}
\end{eqnarray} 
where $\bar{C}={C_{uu}}{T_0}^{-2}$, $\bar{C_1}={(C-P)}{T_0}^{-1}$ 
 for $uu$ component and $\bar{C}={C_{vv}}{T_0}^{-2}$, $\bar{C_1}={(P+C)}{T_0}^{-1}$ for $vv$ component.

The above constitutive relations(\ref{generalTg}-\ref{generalJ}) are new findings. For the chargeless case the relations (\ref{generalTg}-\ref{generalJ}) reproduce the results obtained by the derivative expansion approach 
\cite{Jensen:2012kj}, for the choice $\bar{C}=\frac{\pi}{12}$. Inclusion of charges in that approach requires the consideration of higher order terms which has not been done. We will further elaborate on this in the next section. 

% Before closing the section, let us mention that in order to get a connection between the ($1+1$) dimensional 
%anomalous stress tensor and ideal fluid energy-momentum tensor, we can impose a boundary condition to fix the integration constants. This, as mentioned earlier, corresponds to the Boulware vacuum. 
%This vacuum sets all constants in (\ref{generalTg}),(\ref{generalTU})to zero and yields the structure,
%\begin{equation}
%T_{ab} = T_{ab}^{(g)}+ T_{ab}^{U(1)} =  (\epsilon + {\mathcal{P}})u_au_b + {\mathcal{P}} g_{ab}~,
%\label{ideal}
%\end{equation}
%where the energy density and pressure are identified as,
%\begin{eqnarray}
%&&\epsilon = \frac{1}{24\pi}(2u^a\nabla^b - u^b\nabla^a)\nabla_au_b + \frac{e^2\mu^2}%2\pi}~;
%\nonumber
%\\
%&&{\mathcal{P}} = - \frac{1}{24\pi}u^a\nabla^b\nabla_bu_a +  \frac{e^2\mu^2}{2\pi}~.
%\label{energy}
%\end{eqnarray}
%With the identification (\ref{energy}), the constitutive relation (\ref{ideal}) reproduces the structure for an ideal fluid. Likewise the constitutive relation (\ref{generalJ}) simplifies to,

%\begin{eqnarray}
%%%which also yields the usual ideal fluid structure.

%%%%%%%%%%%%%%%%%%%%%%%%%%%%%%%%%%%%%%%%%%%%%%%%%%%%%%%%%%%%%%%%

\section{\label{chiralsec}Anomalous constitutive relations in chiral hydrodynamics}
In this section, we use the energy momentum tensor and the gauge current to derive  anomalous constitutive relations in chiral hydrodynamics. 
For the chiral gauge theory, as already mentioned, after quantization we have both the trace and diffeomorphism anomaly, where the trace anomaly comes  only from gravity part.

The two-dimensional chiral effective action with $U(1)$ gauge field is given by \cite{Leutwyler:1984nd,Banerjee:2007uc},
\begin{equation}
\Gamma_{(H)} = -\frac{1}{3}z(\omega) + z(A)
\label{totaleffective}
\end{equation}
where
\begin{eqnarray}
&& z(v) = \frac{1}{4\pi}\int d^2xd^2y{\epsilon}^{ab}\partial_a v_b{(x)}\frac{1}{{\Box}}(x,y)\partial_c \left[\left(\epsilon^{cd} + \sqrt{-g}~g^{cd}\right)v_d{(y)}\right]~.
\label{effg}
\end{eqnarray}
%%%%%%%%%%%
The spin connection and the gauge field are denoted, respectively, by $\omega_a$ and $A_a$.
From a variation of the effective action the consistent forms for the energy momentum tensor and the gauge current are obtained. However, we are interested in the covariant forms, so appropriate local polynomials
have to be added. This is possible because energy-momentum tensors and currents are only defined modulo local
polynomials. Then we have
\begin{eqnarray}
\delta \Gamma_H = \int d^2x \sqrt{-g}\left(\frac{1}{2}\delta g_{ab} T^{ab}+ \delta A_a J^a\right) +l,
\label{deltav}
\end{eqnarray} 
where the local polynomial is given by \cite{Leutwyler:1984nd,Banerjee:2007uc},
\begin{eqnarray}
l=\frac{1}{4\pi}\int d^2x \epsilon^{ab}\left(A_a\delta A_b-\frac{1}{3}\omega_a \delta \omega_b -\frac{1}{24}R e^z_a\delta e^z_b\right).
\label{local}
\end{eqnarray} 
Here $e^z_a$ is the zweibein vector which fixes the metric $g_{ab}$.

Including the effect of the local polynomial, the two dimensional covariant chiral energy-momentum tensor can easily be determined by variation of the effective action{(\ref{deltav})} \cite{Banerjee:2007uc}:
\begin{eqnarray}
\label{Tgrav}
&&T_{ab}^{(g)}= \frac{1}{4\pi}\left(\frac{1}{48}D_a \Phi D_b \Phi-\frac{1}{24}D_a D_b\Phi +\frac{1}{24} g_{ab} R \right),
\\
\label{emt}
&&T_{ab}^{U(1)}=\frac{e^2}{4\pi}\left(D_aBD_bB\right),
\end{eqnarray}
%%%%%%%%%%
whereas the covariant chiral gauge current is given by
\begin{equation}
\label{chicurrent}
J^a=-\frac{e^2}{2\pi}D^a B~.
\end{equation}
The total energy-momentum tensor is $T_{ab} = T_{ab}^{(g)} + T_{ab}^{U(1)}$.
The auxiliary fields, $\Phi$ and $B$, are again determined by the solutions of (\ref{boxR}) which are  (\ref{solphi}) and (\ref{solB}).
The chiral nature of $T_{ab}$ and $J_{a}$ are manifested by the presence of the chiral covariant derivative $D_a$ that is defined in terms of usual covariant derivative $\nabla_a$,
\begin{eqnarray}
\label{mag}
~~D_a=\nabla_a-{\bar{\epsilon}}_{ab}\nabla^b =-{\bar{\epsilon}}_{ab}D^b.
\end{eqnarray}
 Based on the above identity it is possible to show the following properties,
 \begin{eqnarray}
 ~~T_{ab}=-\frac{1}{2}\left(\bar{\epsilon}_{ac}{T}^c_b + \bar{\epsilon}_{bc}{T}^c_a\right)+ \frac{1}{2} g_{ab} T^{c}_{c}, \ \ \ J_a= -\bar{\epsilon}_{ab}J^b
\label{TJano}
\end{eqnarray}
which manifest the chiral nature of $T_{ab}$ and  $J_{a}$. 
 
It is easy to check that in null coordinates $D_u = 2\nabla_u$ and $D_v = 0$ and hence this corresponds to the outgoing modes. 
The above stress tensor leads to both trace and diffeomorphism anomalies. The trace anomaly again comes from the gravity part alone: 
\begin{eqnarray}
T^{a(g)}_{a} = \frac{R}{48\pi}; \,\,\,\ T^{a U(1)}_{a} = 0~.
\label{tracec}
\end{eqnarray}
These results are simply obtained by exploiting the chirality condition (\ref{mag}). The stress tensor satisfies  the covariant conservation law, 
\begin{equation}
\nabla_b T^{ab} =  \frac{\bar{\epsilon}^{ab}}{96\pi}\nabla_b R + J_b F^{ab}~,
\label{diffchiral}
\end{equation}
where $F^{ab} = \nabla^a A^b - \nabla^b A^a$ is the gauge field strength. The first term on the right is the covariant diffeomorphism anomaly while the second is the usual Lorentz force term.

Now we will determine the expressions for the components of $T_{ab}$ in null coordinate$~(u,v)~$. Here, as earlier, the ($uv$) components are determined from the trace expression (\ref{tracec}) whereas the other components are found out from (\ref{Tgrav}) and (\ref{emt}) with the use of (\ref{solphi}) and (\ref{solB}):
\begin{eqnarray}
&&T_{uu}^{(g)} = \frac{e^{2\sigma}}{96\pi g_{11}^2} (2\sigma''g_{11} - \sigma'g'_{11}) + C_{uu}~; \,\,\,\ T_{uv}^{(g)} =-\frac{e^{2\sigma}R}{192\pi}~; \,\,\,\ T_{vv}^{(g)}=0~.
\label{uuc}
\\
\label{ab}
&&T^{U(1)}_{uu}=\frac{e^2}{4\pi}\left(P-{A_t}-C\right)^2~; \,\,\,\ 
T^{U(1)}_{vv}=0~; \,\,\,\  T^{U(1)}_{uv}=0.
\end{eqnarray}
Similarly, the components of the current are
\begin{eqnarray}
\label{jmu}
J_u=\frac{e^2}{2\pi}\left(A_t-P+C\right),~~~~~J_v=0.
\end{eqnarray}
Relations given in (\ref{uuc}), (\ref{ab}) yield the expression for the covariant energy momentum tensor with U(1) current  in chiral hydrodynamics.

Finally we deduce the constitutive relations of a chiral fluid in the comoving frame. In this frame the chiral fluid velocity is given in eq.{(\ref{chiralvel})} and fluid variables determined in  eq.{(\ref{fluidvar})}. Therefore, the components can be expressed in the following forms:
 \begin{eqnarray}
\label{chi1}
{T}_{ab}^{(g)} &=& \left[\frac{1}{48\pi}\left(u^c \nabla^d - u^d\nabla^c\right)\nabla_c u_d + e^{-2 \sigma}C_{uu}\right]
\left(2{u}_a{u}_b-{u}_a\tilde{u}_b-\tilde{u}_a{u}_b\right)
\nonumber
\\
&-&\left[\frac{1}{48\pi}\left(u^c \nabla^d \nabla_d u_c\right) -e^{-2\sigma}C_{uu}\right]g_{ab},
\\
\label{chi3}
 {T}_{ab}^{U(1)} &=&\frac{e^2}{4\pi}\left(\mu^2 +C_1 \mu e^{-\sigma} +\frac{{C_1}^2}{4} e^{-2\sigma}\right)
\left(2{u}_a{u}_b-{u}_a\tilde{u}_b-\tilde{u}_a{u}_b+ g_{ab}\right),
\\
\label{chija}
J_a &=&-\frac{e^2}{2\pi}\left(\mu +\frac{{C_1}}{2} e^{-\sigma}\right) u^{(c)}_a.
\end{eqnarray}
%%%%%%%%%%%%%%%%
where, at an intermediate step, we have used the identity,
\begin{eqnarray}
\label{iden}
\tilde{u}_a\tilde{u}_b-{u}_a{u}_b=g_{ab}
\end{eqnarray}
%%%%%%%%%%%
It is now shown that the above constitutive relations can be put in the corresponding forms for an ideal {\it chiral} fluid.

Let us first recall that the constitutive relation for an ideal chiral fluid is different from the  usual expression. To account for the chiral property it is necessary to replace the velocity vectors by chiral velocity vectors \cite{Banerjee:2013qha}. Once this is done the relevant constitutive relation becomes,
\begin{eqnarray}
\label{gen}
{T}_{ab}=\left(\epsilon^c+{\mathcal{P}^c}\right)u_{a}^{(c)}u_{b}^{(c)}+
{\mathcal{P}^c}g_{ab}.
\end{eqnarray}
 where  $u^{(c)}_a$ is defined in (\ref{chiralvel}). The quantities $\epsilon^c$ and $\mathcal{P}^c$ may be regarded as mimicing the energy density and pressure respectively, that appear in the (standard) contribution to the ideal stress tensor,
\begin{eqnarray}
\label{ideal}
{T^0}_{ab}=\left(\epsilon+{\mathcal{P}}\right)u_{a}u_{b}+
{\mathcal{P}}g_{ab}
\end{eqnarray} 
 satisfying $\nabla^a{T^0}_{ab}=0$. The total stress tensor\footnote{The constitutive relations in the previous section have to be interpreted similarly. Expressions (\ref{generalTg}-\ref{generalJ}) yield only the anomalous part. However, contrary to the chiral case, these cannot be expressed in the form (\ref{gen}).}, is a sum of the  contributions from the diffeomorphism invariant part (${T^0}_{ab}$) and the anomalous part ($T_{ab}$),
 \begin{eqnarray}
 \label{ttot}
 T^{(total)}_{ab}={T^0}_{ab}+{T}_{ab}.
 \end{eqnarray}

  It is straightforward to verify the holomorphy condition (\ref{ree}) for (\ref{gen}). Using (\ref{chiralvel}) the ideal chiral constitutive relation (\ref{gen}) is written as, 
 \begin{eqnarray}
 \label{tot}
T_{ab}=\left(\epsilon^c+{\mathcal{P}^c}\right)
\left(2{u}_a{u}_b-{u}_a\tilde{u}_b-\tilde{u}_a{u}_b\right)+\left(\epsilon^c + 2{\mathcal{P}^c}\right)g_{ab}~,
\end{eqnarray}

As done in the last section we introduce the Tolman relation $T=T_0 e^{-\sigma}$ and exploit (\ref{chi1}) and (\ref{chi3}) to write the energy-momentum tensor as,\\
$T_{ab} = T_{ab}^{(g)}+T_{ab}^{U(1)}$
\begin{eqnarray}
\label{chi2}
=\left[\frac{1}{48\pi}\left(u^c \nabla^d - u^d\nabla^c\right)\nabla_c u_d + \bar{C}T^2 +\frac{e^2}{4\pi}\left(\mu^2 +\bar{C_1} \mu T +\frac{\bar{C_1}^2 T^2}{4}\right) \right]
\left(2{u}_a{u}_b-{u}_a\tilde{u}_b-\tilde{u}_a
{u}_b\right)
\nonumber
\\
+ \left[\frac{e^2}{4\pi}\left(\mu^2 +\bar{C_1} \mu T +\frac{\bar{C_1}^2T^2}{4} \right)+ \bar{C}T^2-\frac{1}{48\pi}\left(u^c \nabla^d \nabla_d u_c \right)\right]g_{ab}
\end{eqnarray}
where $\bar{C}=C_{uu}{T_0}^{-2}$ and $\bar{C_1}=2(C-P){T_0}^{-1}$.
Eqn(\ref{chi2}) reproduces (\ref{tot}) with the following identifications:
\begin{eqnarray}
\label{ept}
&&\epsilon^c =  \frac{1}{48\pi}\left(2u^c \nabla^d - u^d\nabla^c\right)\nabla_c u_d + \bar{C}T^2 +\frac{e^2}{4\pi}\left(\mu^2 +\bar{C_1} \mu T +\frac{\bar{C_1}^2T^2}{4} \right);
\\
\label{pt}
&&{\mathcal{P}^c} = -\frac{1}{48\pi}\left(u^c \nabla^d \nabla_c u_d \right).
\end{eqnarray}
Note that in the chiral case, the contribution to $\mathcal{P}^c$ from the gauge field is zero. Also, the constitutive relation (\ref{chija}) for the current is manifestly in the form of an $ideal~ chiral$ fluid relation. Constitutive relations done in this section are general and exact. In the absence of a gauge field the relation (\ref{chi2})  reproduces the result of \cite{Banerjee:2013qha} as well as that found by the gradient expansion method \cite{Jensen:2012kj}.

We note that although $\varepsilon^c$ and $\mathcal{P}^c$ contain derivative/dissipative terms we refer to the relation (\ref{gen}) as `ideal' because it has a structural resemblance with the usual ideal form. The quantities $\varepsilon^c$ and $\mathcal{P}^c$ are not ideal in the sense that they include dissipation.
 %%%%%%%%%%%
\section{\label{comparison}Comparison with derivative expansion approach}

~In this section we make a brief comparison with the derivative expansion approach that is a favoured approach in the context of anomalous hydrodynamics. This would also help in putting our analysis in a proper perspective.

The one-point function of the covariant stress tensor is given by,
\begin{eqnarray}
\label{Tab}
&&T^{ab}=\varepsilon u^a u^b+ \mathcal{P}\tilde{u}^a\tilde{u}^b+\theta\left(\tilde{u}^a u^b+u^a\tilde{u}^b\right)
\end{eqnarray}
which is the general form for a symmetric second rank tensor constructed from the velocity vector $u^a$ 
and its dual $\tilde{u}^a=\bar{\epsilon}^{ab}u_b.$ The explicit values for $\varepsilon$, $\mathcal{P}$ and $\theta$ are provided by the gradient expansion scheme as \cite{Jensen:2012kj},
\begin{eqnarray}
\label{epsipth}
\nonumber
&&\varepsilon = p_{0} T^2 + C_w\left(u^b \nabla^a \nabla_a u_b\right) + 2 C_w\left(u^a\nabla^b-u^b\nabla^a \right)\nabla_a u_b
\\
\nonumber
&&\mathcal{P}=p_{0} T^2 -C_w \left(u^b\nabla^a\nabla_a u_b\right)
\\
&&\theta=C T^2-2 C_g\left(u^a\nabla^b-u^b\nabla^a\right)\nabla_au_b
\end{eqnarray}
where, for simplicity, we consider the charge-less case. The problems of including charge within this scheme will be highlighted later. Here $C_w$ and $C_g$ are the normalisation factors of the conformal(trace) and gravitational(diffeomorphism) anomalies. Likewise, $p_0$ and C are certain response parameters that are undetermined for the moment. In order to determine them it is essential to use earlier results from various (1+1) dimensional conformal field theories. One obtains, 
\begin{eqnarray}\label{pc}
p_0= 4\pi^2 C_w,&& C=-8\pi^2 C_g.
\end{eqnarray}

It is now feasible to deduce the constitutive relation for the stress tensor. Let us first consider the chiral case. Here $C_w=\frac{1}{48\pi}=2C_g$ as seen from (\ref{tracec},\ref{diffchiral}). The response parameters are found to be,
\begin{eqnarray}
\label{piby12}
p_0=\frac{\pi}{12}=-C
\end{eqnarray}
Inserting these values in (\ref{epsipth}) and using the resulting expressions in (\ref{Tab}) reproduces (\ref{chi2}) for $e=0$ (charge-less case) and for the specific choice of $\bar{C}=\frac{\pi}{12}$.

~For the non-chiral case there is no diffeomorphism anomaly (so that $C_g=0$) and there is only a conformal anomaly with $C_w=\frac{1}{24\pi}$ (\ref{trace}). The response parameters (\ref{pc}) are thus given by,
\begin{eqnarray}
\label{piby6}
p_0=\frac{\pi}{6}, && C=0.
\end{eqnarray}
Putting these values in (\ref{epsipth}) and inserting the resulting forms for $\varepsilon, \mathcal{P} $ and $\theta$ in (\ref{Tab}) reproduces the constitutive relation (\ref{generalTg}) for the particular choice $\bar{C}=\frac{\pi}{12}$.

~Some comments are now in order. It is seen that the final constitutive relation can be obtained provided the additional information (\ref{pc}) is known. This is not required in our analysis. Also, the constitutive relation found in the gradient expansion approach corresponds to a specific value $(\bar{C}=\frac{\pi}{12})$ of our results. We may compare this with our approach where the actual value of $\bar{C}$ is left open. 
%This happens because the derivative expansion method is done in a particular frame, the so called landau frame\cite{Jains:2013}. This fixes everything including the particular value of $\bar{C}$. In our approach a frame choice is not necessary so that the results are completely general. Hence there is no disagreement between the derivative expansion approach and the present one. 
Finally, inclusion of charge is quite non-trivial in the gradient expansion approach. The relations (\ref{epsipth}) are no longer exact. There are non-leading (higher derivative) corrections. Likewise, the first relation in (\ref{pc}) also gets modified. These corrections and/or modifications have not been discussed in the literature. Thus the form of the constitutive relation in the presence of both gauge and gravitational anomalies is not clearly spelled out. Hence a one-to-one comparison with our general form is not possible.
 %%%%%%%%%%%%%%%%%%%%%%%%%%%%%%%%%%%%%%%%%%%%%%%%%%%%%%%%%%%%%%%%%%%%%
\section{\label{conclu}Conclusions}

Gauge and gravitational anomalies in two dimensions have played a significant role in different contexts. While such  anomalies can and do occur in various theories, perhaps their most dramatic appearance happens for chiral theories, i.e, where chiral fermions are coupled to the gauge and/or gravitational field. In such theories, due to the lack of a chiral invariant regularisation, the one loop effective action always yields anomalies. Depending upon the regularisation, anomalies may be covariant or consistent. The use of the covariant or consistent structure depends upon the needs of the problem, however it is useful to note that the anomaly vanishing condition is identical in both cases. Since the covariant form preserves the correct transformation property of the anomalous current or stress tensor, it is usually favoured over the consistent form. Indeed, besides their recent influence in hydrodynamics \cite{Sonsuro:2009}-\cite{Banerjee:2013qha}, covariant anomalies have been used to study such diverse phenomena as Hawking effect in black holes \cite{Banerjee:2007qs,Banerjee:2007uc} or the thermal Hall effect in topological insulators \cite{Ston20e:12ud}.

In this paper we have developed a new approach to study the consequences of both gauge and gravitational anomalies in fluid dynamics. 
%This is quite distinct from the usual gradient expansion approach advocated in the literature \cite{Valle:2012em}-\cite{Jains:2013}.
Explicit computations were done in two dimensions leading to the obtention of exact results.
% instead of order by order contributions.
 To highlight the competing role of different anomalies, usual (i.e, nonchiral) and chiral hydrodynamics were treated separately.

Exact closed form expressions for the constitutive relations for the current and stress tensor were obtained. For chiral hydrodynamics these relations could be put in the form of an ideal(chiral) fluid.
% with appropriate identification for the energy density, pressure and chemical potential.
 The constitutive relation for an ideal(chiral) fluid, it may be recalled, is derived from the corresponding relation for an ideal fluid by replacing the velocity vector by the chiral velocity vector. Only then do the current or stress tensor satisfy the special properties of two dimensional chirality \cite{Banerjee:2013qha}. 
%The constitutive relations contain a free parameter that may be fixed by choosing some boundary condition corresponding to the implementation of a specific vacuum state.

In spite of the fact that several papers \cite{Rangamani:2009}-\cite{Banerjee:2013qha} discuss anomalous hydrodynamics including some that are solely devoted to two dimensions \cite{Dubovsky:2011sk}-\cite{Banerjee:2013qha}, the results presented here are both new and from a different approach. The novelty of our approach consists in the inclusion of charge for which no constitutive relation was earlier provided. Since the two dimensional metric is conformally flat, the effective action is exactly known. This leads to exact constitutive relations for either the current or the stress tensor. Chirality imposes additional restrictions that eventually justify the structures for the constitutive relations. In other words not only do we provide new results but we also explain their appearance. However, it would be interesting if our result is compared with the derivative expansion approach by expanding our exact results. But this is beyond the scope of the paper. We leave this exercise for future.

%The gradient expansion approach also yields the constitutive relations found here for the charge less case. However, as shown here, it requires an additional input in the form of equation (\ref{pc}) which simply does not appear in our analysis. Inclusion of charge has not been considered explicitly in the derivative expansion scheme.

%%%%%%%%%%%%%%%%%%%%%%%%%%%%%%%%%
\section*{Acknowledgements}
The research of one of the authors (B.R.Majhi) is supported by a Lady Davis Fellowship at Hebrew University, 
by the I-CORE Program of the Planning and Budgeting Committee and the Israel Science Foundation 
(grant No. 1937/12), as well as by the Israel Science Foundation personal grant No. 24/12.
He is also grateful to the S.N.Bose National Centre for Basic Sciences, India for 
providing necessary facilities.

%%%%%%%%%%%%%%%%%%%%%%%%%%%%%%%%%%%
\begin{appendix}
\section*{\bf{Appendix}}
\section*{\label{Appendix} Solutions for the auxiliary fields}
\renewcommand{\theequation}{A.\arabic{equation}}
\setcounter{equation}{0}  
  The solutions for $\Phi$ and $B$ can be obtained from (\ref{boxR}) in the following way. Let us first concentrate on the solution for $\Phi$.
 Under the background metric (\ref{effectivemetric}), we obtain
\begin{eqnarray}
 &&{{\Box}}\Phi= \frac{1}{\sqrt{-g}}\partial_a \left(\sqrt{-g}~g^{ab}\partial_b\right)\Phi = -e^{-2\sigma} \partial^2_t \Phi + \frac{1}{e^\sigma\sqrt{g_{11}}}\partial_r\left(\frac{e^\sigma }{\sqrt{g_{11}}}\partial_r\right)\Phi;
 \label{boxphi}
 \end{eqnarray}
and $R$ is given by (\ref{Ricci}).  
Now since the metric (\ref{effectivemetric}) is static, it has a timelike Killing vector and so the ansatz for $\Phi$ can be taken as
(\ref{solphi}). Then (\ref{boxphi}) reduces to,
\begin{eqnarray}
\Box\Phi = \frac{1}{e^\sigma\sqrt{g_{11}}} \partial_r\left(\frac{e^\sigma}{\sqrt{g_{11}}}\partial_r\right)\Phi_{0}{(r)}~.
\label{appphi}
\end{eqnarray} 
The Ricci scalar (\ref{Ricci}), multiplied by $e^\sigma\sqrt{g_{11}}$, can be expressed as a total derivative with respective to the radial coordinate $r$:
\begin{equation}
e^\sigma\sqrt{g_{11}} R = -\partial_r\left(\frac{2e^\sigma \sigma^{'}}{\sqrt{g_{11}}}\right)~.
\label{totaldR}
\end{equation}
Hence $\Box\Phi = R$ yields the solution for $\Phi_0(r)$ as given in (\ref{solphi}).

Similarly, the solution for $B$ can be obtained. Substitution of the ansatz in (\ref{solB}) leads to
\begin{eqnarray}
\Box B = \frac{1}{e^\sigma\sqrt{g_{11}}}\partial_r\left(\frac{e^\sigma}{\sqrt{g_{11}}}\partial_r\right)B_{0}~,
\label{appB}
\end{eqnarray}  
whereas $\bar{\epsilon}^{ab}\partial_a A_b = - \frac{1}{e^\sigma\sqrt{g_{11}}} \partial_r A_t$. Hence 
we obtain the solution for $B_0(r)$ as given there.
\end{appendix}

\end{document}